\documentclass[journal=jpcc,manuscript=letter]{achemso}
\usepackage{graphicx}
\usepackage{epstopdf}
\usepackage{tabularx}
\usepackage{graphicx}
\setkeys{acs}{articletitle = true}
\title{Effects of Growth Orientations and Epitaxial Strains on Phase Stability of HfO$_2$ Thin Films }
\author{Shi Liu}
\email{upennliushi@gmail.com;liushi@westlake.edu.cn}
\affiliation{Sensors \& Electron Devices Directorate, U.S. Army Research Laboratory, Adelphi, Maryland 20783, United States.}
\author{Brendan M Hanrahan}
\email{brendan.m.hanrahan.civ@mail.mil}
\affiliation{Sensors \& Electron Devices Directorate, U.S. Army Research Laboratory, Adelphi, Maryland 20783, United States.}

\date{\today}
\begin{document}
\begin{abstract}{
The discovery of ferroelectricity in both pure and doped HfO$_2$-based thin films have revitalized the interest in using ferroelectrics for nanoscale device applications. To take advantage of this silicon-compatible ferroelectric, fundamental questions such as the origin of ferroelectricity and better approach to controlled realization of ferroelectricity at the nanoscale need to be addressed. The emergence of robust polarization in HfO$_2$-based thin films is considered as the cumulative effect of various extrinsic factors such as finite-size effects and surface/interface effects of small grains, compressive stress, dopants, oxygen vacancies, and electric fields. The kinetic effects of phase transitions and their potential impacts on the emergence of ferroelectricity in HfO$_2$ at the nanoscale are not well understood. In this work, we construct the transition paths between different polymorphs of hafnia with density functional theory calculations and variable-cell nudged elastic band technique. We find that the transition barriers depend strongly on the mechanical boundary conditions and the transition from the tetragonal phase to the polar orthorhombic phase is a fast process kinetically under clamping. The effects of growth orientations and epitaxial strains on the relative stability of different phases of HfO$_2$ are investigated. The two orthorhombic phases, polar $Pca2_1$ and non-polar $Pbca$, become thermodynamically stable in (111)-oriented thin films over a wide range of epitaxial strain conditions. This work suggests a potential avenue to better stabilize the ferroelectric phase in HfO$_2$ thin films through substrate orientation engineering.
}
\end{abstract}
\maketitle
\newpage
\section{Introductions}
Ferroelectrics characterized by the switchable spontaneous polarization have long been considered as a candidate material to realize low-power high-speed nonvolatile memories and logic devices.~\cite{Buck52MT,Ross57patent,Scott07p954} However, the difficulty of integrating perovskite ferroelectrics into complementary metal-oxide-semiconductor (CMOS) processes~\cite{Pinnow04pK13} has hindered device scaling to the sub-100 nm regime.~\cite{McAdams04p667} The discovery of ferroelectricity in both pure and doped HfO$_2$ thin films~\cite{Boscke11p102903,Mller11p114113,Mueller12p2412,Schroeder14p08LE02,Park15p1811,Pal17p022903} have revitalized the interest in using ferroelectrics in nanoscale devices~\cite{Pei17p1236} because HfO$_2$ is a well-studied CMOS-compatible gate dielectric that is thermodynamically stable on silicon.~\cite{Gutowski02pB3.2} Moreover, contrary to conventional perovskite-based ferroelectric thin films where the depolarization field due to an incomplete screening of surface charges often leads to the instability of out-of-plane polarization,~\cite{Batra73p3257,Wurfel73p5126,Ma02p386} HfO$_2$-based thin films possess robust electrical polarization at the nanoscale, ideal for device miniaturization.~\cite{Boscke11p102903} It is suggested that HfO$_2$ may well be a new type of ferroelectric materials that only exhibits spontaneous polarization at the nanometer length scale but not in the bulk form.~\cite{Wei18}

The observation of ferroelectricity in HfO$_2$ is unexpected, given that this simple binary oxide has been studied extensively for decades as high-$\kappa$ dielectrics.~\cite{Robertson05p327} Bulk HfO$_2$ adopts the monoclinic (M) $P2_1/c$ phase at the room temperature and pressure, and transforms to the tetragonal (T) $P4_2/nmc$ phase and the cubic $Fm\bar{3}m$ phase with increasing temperature, and becomes the ``antiferroelectric-like" orthorhombic (AO) $Pbca$ phase at a higher pressure. All these phases are centrosymmetric and thus non-polar. B\"oscke {\em et al.} first reported the presence of ferroelectricity in thin films of SiO$_2$-doped hafnium oxide with a thickness of 10 nm.~\cite{Boscke11p102903} Since then, many divalent and trivalent dopants were found to be able to induce ferroelectricity in hafnia thin films.~\cite{Mller11p114113,Mueller12p2412,Schroeder14p08LE02,Starschich17p333} Robust polarization of 10$\mu$C/cm$^2$ was observed even in dopant-free hafnia thin films in a thickness range of 4-12 nm.~\cite{Polakowski15p232905} 

The origin of ferroelectricity in HfO$_2$-based thin films has attracted intensive studies in recent years. Combined experimental and theoretical studies eventually lead to the conclusion that the polar orthorhombic (PO) $Pca2_1$ phase is most likely responsible for the ferroelectricity.~\cite{Park15p1811, Huan14p064111,Sebastian14p140103,Sang15p162905,Materlik15p134109} Based on density functional theory (DFT) calculations, the M phase is the most stable phase, followed by the polar PO phase and the non-polar T phase.~\cite{Huan14p064111,Sebastian14p140103,Materlik15p134109} The stabilization of the metastable PO phase in thin films is attributed to a variety of extrinsic factors such as finite-size effects and surface/interface effects associated with small grain size,~\cite{Materlik15p134109, Park15p1811, Polakowski15p232905, Batra16p172902,Knneth17p205304,Park17p9973} strains of different origins,~\cite{Shiraishi16p262904,Batra17p4139} dopants,~\cite{Schroeder14p08LE02, Starschich17p333,Park17p4677,Xu17p124104,Batra17p9102} oxygen vacancies,~\cite{Xu16p091501,Pal17p022903} and electric fields.~\cite{Batra17p4139} 

In the phenomenological surface energy model developed by Materlik {\em et al.} that takes into account the surface energy contribution to the phase stability, the hafnia grains of nanometer size ($<4$~nm) will favor the ferroelectric phase because of the lower surface energy of the PO phase than that of the M phase.~\cite{Materlik15p134109} Batra {\em et al.} evaluated the surface energies of different phases of hafnia with DFT and found that the PO phase has lower (001) surface energy but higher (100) and (010) surface energies than those of the M phase.~\cite{Batra16p172902} They also found that the surface energies of the T phase are consistently lower than the M phase for all surface planes, suggesting a surface-induced formation of the non-polar T phase at the nanoscale followed by a transition to the ferroelectric PO phase due to other extrinsic perturbations.~\cite{Batra16p172902} 

The epitaxial compressive strain is considered to play an important role in inducing the ferroelectric PO phase.~\cite{Sebastian14p140103,Batra17p4139} However, DFT calculations have shown that neither hydrostatic pressure nor biaxial compressive stress alone is enough to make the polar phase the most stable phase. Therefore, the application of external electric fields is required to drive the non-polar to polar phase transition, ~\cite{Batra17p4139} which may explain the ``wake-up effect" in hafnia thin films.~\cite{Zhou13p192904,Schroeder14p08LE02, Schenk14p041103} The presence of oxygen vacancies is often detrimental to the device functionality by causing leakage current, aging, and fatigue in ferroelectrics, but is proposed to be beneficial for realizing the ferroelectric PO phase.~\cite{Park17p9973} Recent experiments demonstrated an enhanced ferroelectricity in sub-10~nm dopant-free hafnia thin films by lowering the oxidant dose (thus increasing the oxygen vacancy concentration) during growth,~\cite{Park17p9973} consistent with first-principles results that ionized oxygen vacancies can stabilize the metastable PO phases.~\cite{Lee08p012102} It appears that the emergence of ferroelectricity in HfO$_2$ thin films is indeed the cumulative effect of various extrinsic factors. 

Previous studies on epitaxial strain engineering of ferroelectricity have mostly focused on thin films grown on $(00l)$-oriented substrates.~\cite{Schlom07p589} There is now growing interest in applying biaxial strain along other crystallographic planes such as (101) and (111) by growing thin films on substrates of different cuts.~\cite{Damodaran16p263001} The substate orientation offers an unconventional route to tune the mechanical boundary conditions that can substantially affect the phase stability and ferroelectric domain morphologies. For example, DFT calculations suggest that (111)-oriented BaTiO$_3$ thin films have  strain-driven phase transitions drastically different from the (001)-oriented thin films, and (110)-oriented BaTiO$_3$ may support three distinct monoclinic phases that are not present in bulk BaTiO$_3$.~\cite{Angsten17p174110} Experimental works on (111)-oriented PbZr$_x$Ti$_{1-x}$O$_3$ thin films revealed a high-density, nanotwinned domain morphology consisting of degenerate polarization variants that enable both direct 180$^\circ$ and multi-step 90$^\circ$ switching.~\cite{Xu14p3120,Xu15p79,Xu19p1}  In this regard, it is important to explore the effect of substrate orientation on the ferroelectric properties of HfO$_2$.~\cite{Park14p072901} Recently, it was discovered that Hf$_{0.5}$Zr$_{0.5}$O$_2$ thin films grown on (001)-oriented  La$_{0.7}$Sr$_{0.3}$MnO$_3$/SrTiO$_3$ substrates predominantly adopt the (111) orientation and display large polarization values up to 34$\mu$C/cm$^2$,~\cite{Wei18} highlighting the potential impact of crystal orientations on the functional properties of HfO$_2$-based thin films. Moreover, HfO$_2$-based films are often polycrystalline consisting of grains of different orientations such that only grains with the polar axis along the out-of-plane direction will contribute to the out-of-plane polarization in (001)-oriented films. Growing unconventionally oriented films such as (111)-oriented films will guarantee stable out-of-plane polarization, which is beneficial for the usage of polycrystalline films for high density integration via downscaling of the lateral dimensions.~\cite{Katayama16p112901}
\raggedbottom

In this work, we start by examining the ideal transition barriers between different phases of HfO$_2$ with first-principles methods, aiming to reveal the underlying multi-well potential energy surface that is critical for the understanding of the kinetic effects of phase transitions. We find that the three low-energy phases, $P2_1/c$, $Pca2_1$, and $Pbca$, are separated by significant barriers ( $> 0.1$~eV/f.u.), and the high-temperature tetragonal $P4_2/nmc$ phase serves as an important intermediate structure bridging different phases. Moreover, the transformation from the tetragonal phase to the polar phase is kinetically fast, and tuning the mechanical boundary conditions can make it the most favorable transition path. Growing unconventionally oriented crystals on substrates of different cuts provides an additional knob to control the mechanical boundary conditions. We evaluate the energies of different polymorphs of hafnia with biaxial deformations applied in the \{100\}, \{110\}, and \{111\} planes. Our calculations reveal that the two orthorhombic phases, polar $Pca2_1$ and non-polar $Pbca$, become more stable than the monoclinic phase in (111)-oriented films over a range of epitaxial strain conditions, suggesting a new route to better stabilize the polar phase in HfO$_2$ thin films through substrate orientation engineering.

 \section{Computational Methods}
{\bf {\em ab initio} studies on the phase transition paths }

We use the variable-cell nudged elastic band (VC-NEB) technique~\cite{Qian13p2111} implemented in the USPEX code~\cite{Oganov06p244704,Lyakhov13p1172,Oganov11p227} to study the transition paths between different polymorphs of hafnium oxides. Specifically, we investigate following solid-solid phase transitions: T $\leftrightarrow$ PO,   T $\leftrightarrow$ AO,  T $\leftrightarrow$ M, and  AO $\leftrightarrow$ PO. Each transition path is constructed by 40 images. The {\em ab initio} calculations are carried out using local density approximation implemented in Quantum Espresso~\cite{Giannozzi09p395502} with ultrasoft pseudopotentials from the Garrity, Bennett, Rabe, Vanderbilt high-throughput pseudopotential set.~\cite{Garrity14p446} A plane-wave cutoff of 50 Ry, a charge density cutoff of 250 Ry, and a reciprocal-space resolution of 0.26 ($\rm \AA^{-1}$) for $k$-points generation are used. We choose 0.025 eV/$\rm \AA$ for the root-mean-square (RMS) forces on images as the halting criteria condition for VC-NEB calculations. The variable elastic constant scheme~\cite{Henkelman00p9901} is utilized by setting the spring constant between the neighboring images in the range of $2.0-6.0$ eV/$\rm \AA^2$. 

The performance of NEB calculations requires a predefined atomic mapping scheme upon which the initial transition path consisting of a discreet set of configurations can be constructed, typically by a linear interpolation of the Cartesian coordinates between the initial and final states. VC-NEB introduces additional lattice degrees of freedom.~\cite{Qian13p2111} Therefore, there could be multiple transition paths connecting two phases depending on the choice of atomic and crystal axis (orientation) mapping (See details in the Supporting Information).  Here we only report the minimum energy path connecting two phases. 

{\bf {\em ab initio} studies on the effects of crystal growth orientations}

\begin{figure}[!]
\centering
\includegraphics[scale=1.0]{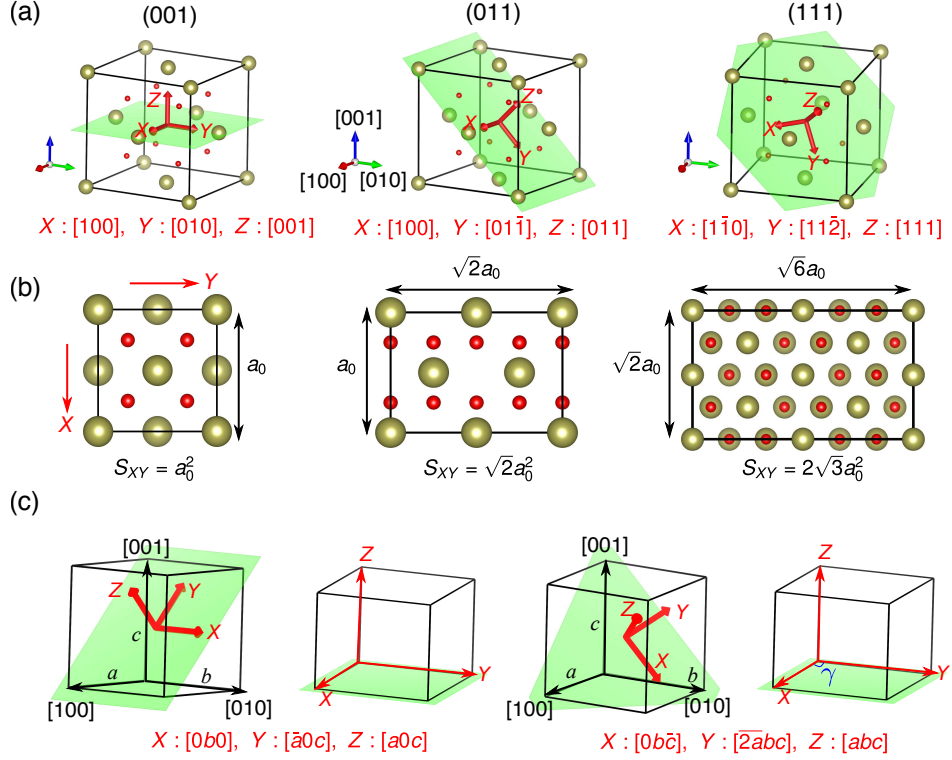}\\
 \caption{(a) Orientation measurement axes ($X$,$Y$,$Z$) for the (100)-, (011)-, and (111)-growth orientations displayed in a cubic HfO$_2$ structure. Cubic crystallographic directions are indicated in square brackets. Hf and O atoms are denoted by golden and red spheres, respectively. (b) Schematic atomic arrangements for the (100), (011), and (111) planes of cubic HfO$_2$ viewed along the $Z$ axis. $a_0 = 5.00~\rm{\AA}$ is the LDA lattice constant optimized with LDA GBRV pseudopotentials. $S_{XY}$ is the $XY$ planar area. (c) Illustrations of the (101) and (111) orientations denoted as $X[0b0]Y[\bar{a}0c]Z[a0c]$ and $X[0b\bar{c}]Y[\overline{2a}bc]Z[abc]$ for a non-cubic structure. The $XY$ plane is the epitaxial matching plane where the lattice vectors are scaled and fixed when applying biaxial deformations while the $Z$ axis is chosen as the out-of-plane direction and is optimized along with the atomic positions. }
  \label{setup}
 \end{figure}

We explore three biaxial strain states corresponding to the epitaxial growth of (100)-, (110)-, and (111)-oriented films. The following four phases are considered: M, T, PO, and AO. For the cubic HfO$_2$ structure, the measurement axes distinguishing different growth orientations are illustrated in Figure~\ref{setup}a: the (001)-oriented film has [100] along $X$, [010] along $Y$, [001] along $Z$; similarly, the (011)-oriented film has [100] along $X$, $[01\bar1]$ along $Y$, and [011] along $Z$; the [111]-oriented system has $[1\bar10]$ along $X$, $[11\bar2]$ along $Y$, and [111] along $Z$. The $XY$ plane is chosen as the epitaxial matching plane and the $Z$ axis is considered as the out-of-plane direction of a thin film. Figure~\ref{setup}b shows the atomic arrangements viewed along the $Z$ axis, where $a_0 = 5.00~\rm{\AA}$ is the optimized lattice constant of cubic HfO$_2$. Crystal structures were visualized using VESTA.~\cite{Momma11p1272}

Because the four phases of HfO$_2$ have symmetry lower than the cubic phase, there are three different \{001\} planes for the M, PO, and AO phases and two for the T phase, three \{101\} planes and three \{111\} planes for all four phases, respectively. Considering a unit cell with lattice constants $a \ne b \ne c$, we introduce following notations to represent different orientations: the three \{001\} planes are labeled as $X[a00]Y[0b0]Z[00c]$, $X[0a0]Y[0c0]Z[00b]$, and $X[0b0]Y[0c0]Z[00a]$; the three \{101\} planes are $X[a00]Y[0b\bar{c}]Z[0bc]$, $X[0b0]Y[\bar{a}0c]Z[a0c]$, and $X[00c]Y[a\bar{b}0]Z[ab0]$; the three \{111\} planes are $X[a\bar{b}0]Y[ab\overline{2c}]Z[abc]$, $X[a0\bar{c}]Y[a\overline{2b}c]Z[abc]$, and $X[0b\bar{c}]Y[\overline{2a}bc]Z[abc]$. Here we assume the short axis $a$ of the unit cell is along [100] and the long axis $c$ is along [001]. As an example, the notation $X[0b0]Y[\bar{a}0c]Z[a0c]$ means the $b$ axis is along $X$, $[\bar{a}0c]$ is along $Y$, and $[a0c]$ is along $Z$. Similarly, $X[0b\bar{c}]Y[\overline{2a}bc]Z[abc]$ represents a crystal orientation with $[0b\bar{c}]$ along $X$, $[\overline{2a}bc]$ along $Y$, and $[abc]$ along $X$ (see graphical illustrations in Figure~\ref{setup}c). It is noted that because the unit cell of AO phase has 24 atoms and the lattice constant along the [100] direction is 2$a$, its [022] and [122] directions coincide with the [011] and [111] directions of the 12-atom unit cell of M and PO phases. In following discussions we will ignore this subtle difference and will only use notations based on the 12-atom unit cell. 

For each phase, we first construct a supercell with the prescribed crystal orientation and then fully relax the lattice vectors and atomic positions. After the equilibrium lattice parameters are obtained, the biaxial strain is applied in $XY$ plane by scaling the lattice vectors along $X$ and $Y$ while conserving the in-plane lattice angle $\gamma$ (Figure~\ref{setup}c). We make a special note here that the biaxial strain states studied in this way do not consider the effect of in-plane shear strain as the value of $\gamma$ is fixed. The internal coordinates are fully optimized while allowing the length and direction of the $Z$ axis to relax until the forces on the atoms are less than $1.0\times10^{-4}$ Ry/Bohr. 

 \section{Results and Discussions}
\begin{figure}[!]
\centering
\includegraphics[scale=1.0]{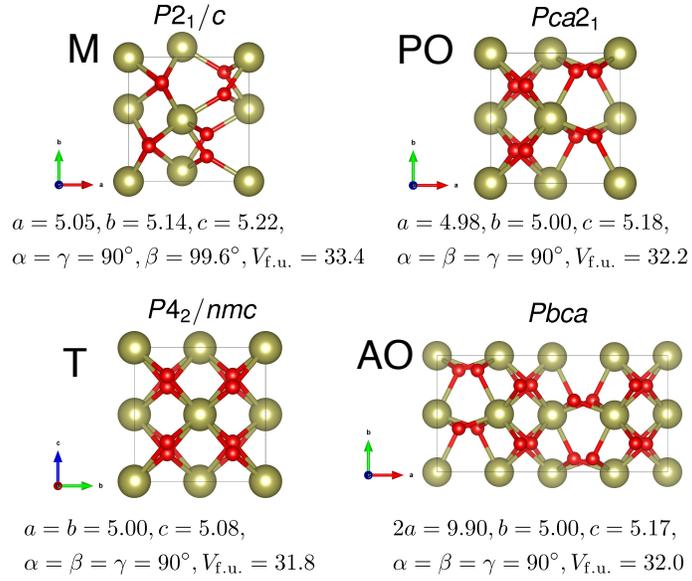}\\
 \caption{Optimized lattice parameters of the $P2_1/c$ (M), $Pca2_1$ (PO), $P4_2/nmc$ (T), and $Pbca$ (AO) phases of HfO$_2$. Lattice constants in the unit of $\rm \AA$ and volume per formula unit ($V_{\rm f.u.}$) in the unit of $\rm \AA^3$. }
  \label{lattice}
 \end{figure}

\begin{figure}[b]
\centering
\includegraphics[scale=1.0]{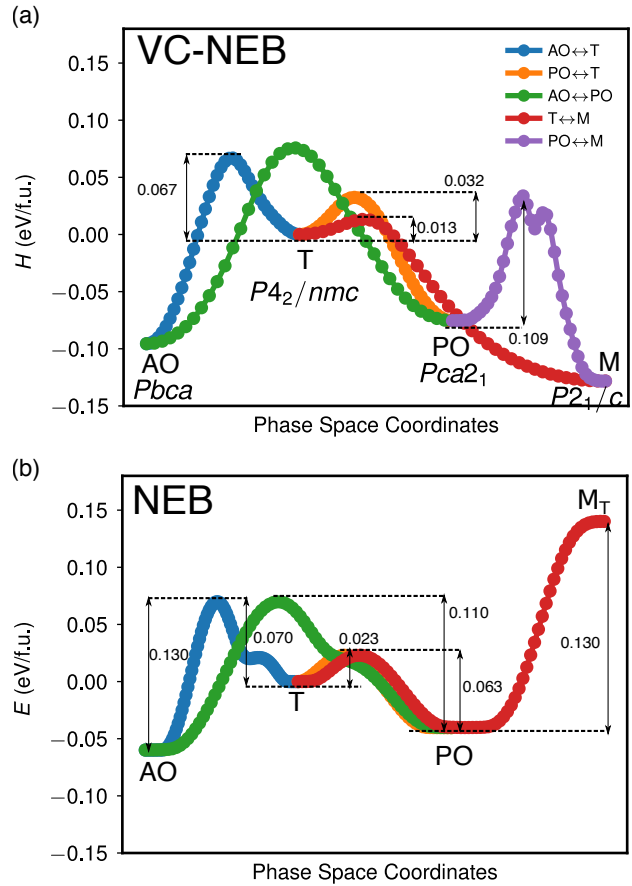}\\
 \caption{(a) Minimum energy paths connecting different phases of HfO$_2$ obtained with VC-NEB. (b) Phase transition paths obtained with clamped-cell NEB by fixing lattice constants to the values of the T phase. Transition barriers in the unit of eV/f.u. are labeled for selected paths.} 
  \label{neb}
 \end{figure}

{\bf Phase Transitions of HfO$_2$ polymorphs}
Figure~\ref{lattice} reports the optimized lattice parameters of different polymorphs using LDA. Our results are consistent with previous DFT studies~\cite{Huan14p064111,Sebastian14p140103,Materlik15p134109}. 
In reference to the T phase, the energy per formula unit (f.u.) of M, AO, and PO phases is $-128$, $-95$, and $-75$ meV/f.u., respectively. Figure~\ref{neb}a shows the minimum energy paths identified with VC-NEB. The transition barriers are used to gauge the kinetics and are reported in Table 1. We find that the three phases, M, AO, and PO, are separated by large barriers ($>100$~meV/f.u. $\approx 1160$~K).  In contrary, the transition from the T phase to the other three phases only needs to overcome a small enthalpy barrier, $\Delta H^{\neq}$(T$\rightarrow$AO) = 67, $\Delta H^{\neq}$(T$\rightarrow$ PO) = 32$,\Delta H^{\neq}$(T$\rightarrow$ M) = 13 meV/f.u., respectively. We propose the T phase is likely to serve as the ``precursor" phase during the phase transitions between M, AO, and PO. Notably, the transformation from the T phase to the ferroelectric PO phase is faster {\em kinetically} than to the non-polar orthorhombic AO phase despite the AO phase being lower in enthalpy. This implies the formation of the ferroelectric PO phase is feasible due to the relatively low kinetic barrier. 

We now compare the two phase transitions that have low barriers: T$\rightarrow$ M and T$\rightarrow$ PO. It is found that the transition T$\rightarrow$ M involves a large change (11\% increase) in lattice angle $\beta$ (between $a$ and  $c$ axes) whereas the transition T$\rightarrow$ PO undergoes a modest unit cell variation (see details in Supporting Information). The substantial shear deformation required to realize the T$\rightarrow$ M transition hints at a stronger impact of substrate clamping. To test this hypothesis, we carry out conventional clamped-cell NEB calculations where the lattice constants are fixed to the values of the T phase, assuming the T phase is the ``precursor" phase during the growth of thin films.~\cite{Park18p716} As revealed in Figure~\ref{neb}b, constraining the lattice degrees of freedom drastically affects the transition paths. First, the ``artificial'' M phase with the same lattice constants of the T phase, denoted as M$_{\rm T}$, becomes a high energy phase. This M$_{\rm T}$ phase experiences highly anisotropic stresses, $\sigma_{13} = \sigma_{31} = -10.8$~GPa and $\frac{1}{3}$tr($\sigma_{ij}$) = 1.37~GPa. The energy of the non-polar AO phase remains lower than that of the polar PO phase. These results are in line with previous studies that a compressive pressure tends to destabilize the M phase though never makes the PO phase the lowest-energy phase. Second, the transition T$\rightarrow$ M$_{\rm T}$ involves the ferroelectric PO phase as the intermediate structure, in agreement with the dopant-induced phase transition route (M$\rightarrow$O$\rightarrow$T) observed experimentally.~\cite{Park17p4677,Xu17p124104}  Finally, we find that the transition T$\rightarrow$ AO is kinetically slower than the transition T$\rightarrow$ PO, and AO and PO phases are separated by a large barrier, making the direct transition difficult. Therefore, the transition T$\rightarrow$ AO is kinetically favored over all the other transition paths investigated here in a clamped cell. 

The VC-NEB and clamped-cell NEB investigations highlight the kinetic and strain effects on the emergence of ferroelectric phase in HfO$_2$ thin films. Specifically, the barrier heights estimated with the VC-NEB method are more relevant to phase transitions in bulk-like environments or films with relaxed strain, whereas the barrier heights obtained with the clamped-cell NEB method reflect the kinetics of phase transitions in strained grains or thin films. Our results show that when the shear strain required for the T$\rightarrow$ M transition is inaccessible, the T phase is more likely to become the ferroelectric PO phase {\em kinetically} even though the non-polar AO phase is still favored {\em thermodynamically}. In this regard, growing HfO$_2$-based films on unconventionally oriented substrates may introduce appropriate mechanical boundary conditions that enhance the ferroelectricity, thus inspiring our following investigations. 
  
\begin{figure}[b]
\centering
\includegraphics[scale=1.0]{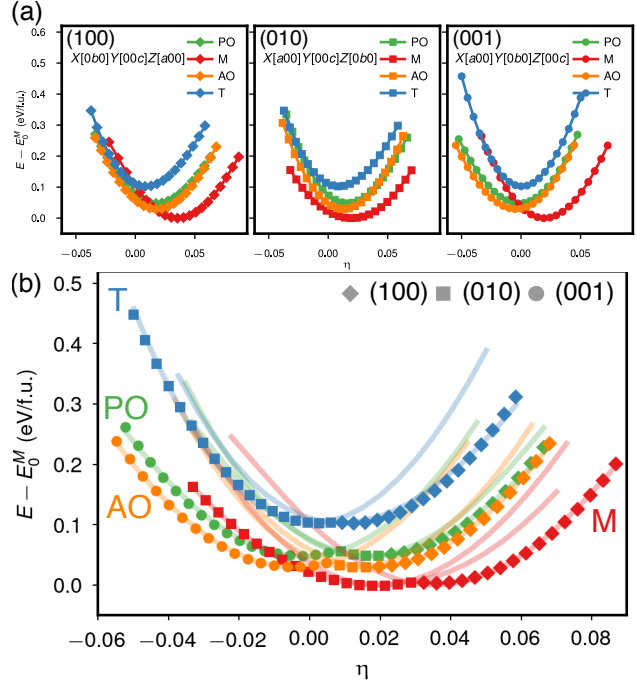}\\
 \caption{(a) Energies of polymorphs of HfO$_2$ in response to epitaxial strains ($\eta$) in \{001\} planes. The energy of the equilibrium bulk M phase ($E_0^M$) is chosen as the reference. Epitaxial strain is calculated with respect to cubic HfO$_2$ as described in the text. (b) Energy vs strain diagram of polymorphs of HfO$_2$ of (001), (010), and (001) orientations. The M, T, AO, and PO are colored in red, blue, orange, and green, respectively. For a given strain and phase, the lowest-energy orientation is highlight with filled diamond for (100), square for (010), and circle for (001).}
  \label{001}
 \end{figure}

{\bf Effects of Crystal Orientations}  We first study the conventional (001)-oriented films. The effect of epitaxial strain in the (001) plane ($X[a00]Y [0b0]Z[00c]$) on the relative stability of different phases of hafnia was already explored in ref.~\citenum{Batra17p4139} with DFT using PBE exchange-correlation functional.~\cite{Perdew96p3865,Perdew97p1396} Here we systematically investigate the effects of biaxial deformations in all \{001\} planes on the energetics of M, T, AO, and PO phases with LDA. Figure~\ref{001} shows the energy as a function of epitaxial strain defined as $\eta = (a-a_0)/a_0$ where $a_0$ is the optimized lattice constant of cubic HfO$_2$ and $a$ is the effective lattice constant estimated from the $XY$ planar area, $a=\sqrt{S_{XY}}$ . We first compare the relative phase stability due to the epitaxial strains in (100), (010), and (001) planes separately (Figure~\ref{001} top). The trend in phase stability of (001)-oriented films revealed by our LDA calculations agrees with the PBE results: the compressive strain tends to stabilize the two orthorhombic phases AO and PO over the M phase, while the energy of the non-polar AO phase is always lower than the polar PO phase.~\cite{Batra17p4139} Similar trend is also found for equibiaxial deformations imposed in the (100) plane. However, the M phase remains the lowest-energy phase in (010)-oriented films. 

To distinguish different crystal orientations, we introduced the convention that the $a$, $b$, and $c$ axes of the unit cell are aligned along the Cubic crystallographic directions [100], [010], and [001], respectively (Figure~\ref{setup}). However, it is possible that the $c$ axis of the crystal will align along the [100] or [010] direction of the substrate during the growth or the cooling process after film deposition,~\cite{Katayama16p134101} leading to (100)- or (010)-oriented crystal grown on (001)-oriented substrates. For this reason, it is important to compare energies of all differently-oriented crystals (Figure~\ref{001} bottom). We find that the compressive strain ($\eta < 0$) induces a M $\rightarrow$ AO transition above a critical strain $\eta \approx -0.4\%$, whereas the ferroelectric PO phase never becomes the lowest-energy phase. 

\begin{figure}[b]
\centering
\includegraphics[scale=1.0]{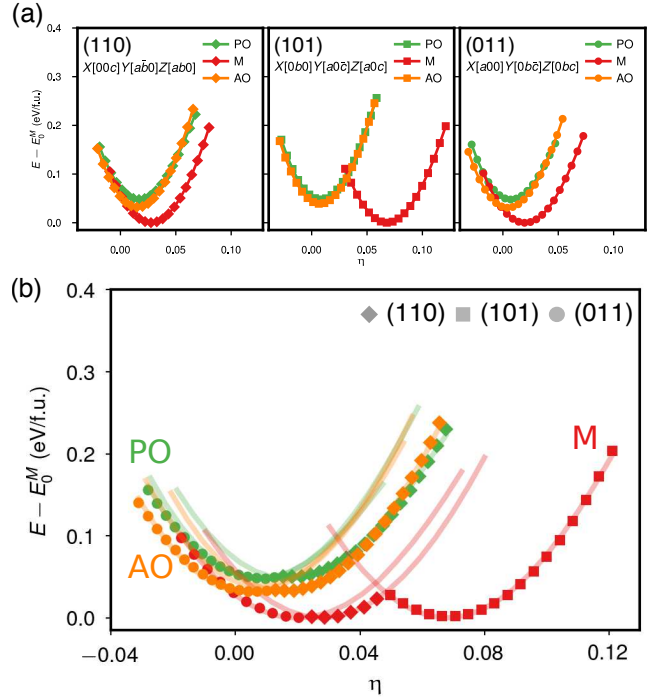}\\
 \caption{Energies of polymorphs of HfO$_2$ in response to epitaxial strains ($\eta$) in \{101\} planes. Epitaxial strain is calculated with respect to cubic HfO$_2$ as described in the text. (b) Energy vs strain diagram of polymorphs of HfO$_2$ of (110), (101), and (011) orientations. The M, AO, and PO are colored in red, orange, and green, respectively. For a given strain and phase, the lowest-energy orientation is highlight with filled diamond for (110), square for (101), and circle for (011).}
  \label{101}
 \end{figure}

Next, we consider the effect of biaxial strain applied in \{101\} planes on the relative stability of M, AO, and PO phases (Figure~\ref{101}). The epitaxial strain is measured with respect to the cubic HfO$_2$, $\eta = (a-a_0)/a_0$, where the effective in-plane lattice constant is defined as $a = \sqrt{S_{XY}/\sqrt{2}}$ (see Figure~\ref{setup}). In (101)-oriented films, we find the {\em optimal equilibrium strain} (at which the energy is a local minimum) of the two orthorhombic phases ($\eta \approx 1\%$) is well separated from that of the M phase ($\eta \approx 7\%$). This suggests it is possible to stabilize AO and PO phases {\em thermodynamically} by depositing HfO$_2$ on a substate with an in-plane lattice constant of $\approx$ 5.05~${\rm\AA}$ if the $[a0c]$ lattice vector  of the crystal is controllably oriented along the out-of-plane direction. However, when the constraint on the crystal orientation is lifted, the M phase is favored over a wide range of epitaxial strain conditions (Figure~\ref{101} bottom). There is also a compressive strain-induced M $\rightarrow$ AO transition at $\eta \approx -0.3\%$.

\begin{figure}[b]
\centering
\includegraphics[scale=1.0]{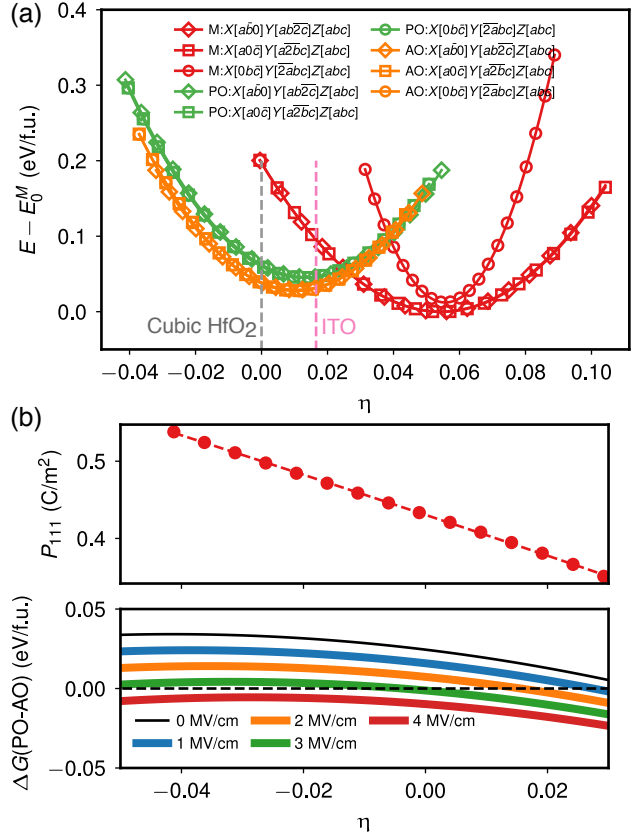}\\
 \caption{(a) Energies of polymorphs of HfO$_2$ in response to equibiaxial deformations applied in \{111\} planes. Epitaxial strain is calculated with respect to cubic HfO$_2$ (gray dashed line) as described in the text. The strain corresponding to Sn-doped In$_2$O$_3$ (ITO)~\cite{Katayama16p134101} is highlighted as well. (b) 
Strain dependence of $P_{[111]}$ and the free energy difference between PO and AO phases ($\Delta G$(PO-AO)) in response to an electric field along [111].}
 \label{111}
 \end{figure}

We now focus on (111)-oriented films where the effective in-plane lattice constant is calculated using $a = \sqrt{S_{XY}/2\sqrt{3}}$ (see Figure~\ref{setup}b). The computed epitaxial strain diagram as shown in Figure~\ref{111} reveals a feature unique to (111)-oriented films. In all (111)-oriented films, both AO and PO phases have energy lower than the M phase at the optimal equilibrium strain ($\eta \approx 1\%$ for AO and $\eta \approx 1.5\%$ for PO) characterized by a local energy minimum. In comparison, the M phase remains the lowest-energy phase at the optimal equilibrium strains of PO and AO phases in (100)/(101)-oriented thin films. In this regard, it is more likely to realize the orthorhombic phase thermodynamically by growing (111)-oriented thin films on selective substrates that offer the right in-plane strains even in the absence of electric field.  

The remaining issue is the consistent higher energy of the ferroelectric PO phase relative to the ``antiferroelectric-like" AO phase. We propose two mechanisms potentially responsible for the formation of ferroelectric PO phase in thin films. Kinetically, as revealed by our VC-NEB calculations, the transition from the T phase to the PO phase is faster than to the AO phase in a bulk-like environment. The substrate orientation and the underlying mechanical clamping will likely affect the transition barriers and kinetics. We suggest the intrinsic barrier heights estimated with VC-NEB are good approximations to those in (111)-oriented thin films at the optimal equilibrium strain where the crystal is in a nearly stress-free equilibrium state similar to a bulk condition. Therefore, if the T phase forms first during the high temperature annealing process when preparing the thin films, a significant portion of the T phase may transform to the PO phase because of the low activation energy of the T $\rightarrow$ PO transition.  This kinetic argument is supported by recent experimental work on the effect of heat treatment on the formation of ferroelectric epitaxial (111)-oriented 7\% Y-doped HfO$_2$ (YHO7) thin films.~\cite{Mimura19pSBBB09} In this work, the as-deposited YHO7 films were heat-treated by a rapid thermal furnace at various temperatures. It was found that the as-deposited and lower temperature treatment resulted in M phase whereas a higher temperature treatment leads to more O/T phases. Further in situ high-temperature X-ray diffraction measurement indicates that the M phase has transformed to the tetragonal phase above 950~$^\circ$C, and the O phase appears below 300~$^\circ$C during the cooling process. Our DFT calculations therefore offer a possible explanation to the origin of ferroelectricity in (111)-oriented YHO7 thin films consisted of single orthorhombic phase.~\cite{Mimura16p052903} 

Thermodynamically, the presence of an external field along the polar axis will favor the PO phase and induce the AO $\rightarrow$ PO transition, which could be one factor responsible for the ``wake-up" effect observed experimentally in many HfO$_2$-based thin films.~\cite{Park16p15466,Starschich17p333} We estimate the free energy $G$ under an electric field along the out-of-plane [111] direction ($\mathcal{E}_{111}$), $G = E_{\rm DFT} -  \mathcal{E}_{111}P_{[111]}V_{\rm f.u.}$, where $E_{\rm DFT}$ is the DFT energy per formula unit and $P_{[111]}$ is the polarization.  The strain dependence of $P_{[111]}$ and the free energy difference between PO and AO phases ($\Delta G$(PO-AO)) are shown in Figure~\ref{111}b.  An electric field of $\approx$2 MV/cm is already high enough to drive the  AO $\rightarrow$ PO transition at some strain condition, consisent with experimental results obtained in (001)-oriented films.~\cite{Park16p15466,Starschich17p333} We make a special note that the theoretical switching field estimated here assumes a homogenous bulk transition and could be much higher than the physical switching field arising from the ``nucleation-and-growth" mechanism at the domain wall.~\cite{Shin07p881,Liu16p360} Further studies are required to quantify the effects of growth orientation and epitaxial strains on the magnitude of switching field relevant to the ``wake-up" effect.

 \section{Conclusions}
In summary, we have investigated the kinetic effects of phase transitions on the emergence of ferroelectricity in HfO$_2$ thin films by quantifying the transition barriers between different polymorphs of hafnia with density functional theory calculations. The multi-well potential energy surface obtained with the variable-cell nudged elastic band technique suggests that the mixture of different phases often presented in HfO$_2$-based thin films has both kinetic and thermodynamic reasons. We propose the tetragonal $P4_2/nmc$ phase, likely formed during the annealing process at high temperatures, is the key ``precursor" phase, responsible for the formation of monoclinic and orthorhombic phases because of the low transition barriers. One important finding is that the transition from the tetragonal phase to the polar orthorhombic $Pca2_1$ phase is kinetically faster than the transition to the competing non-polar orthorhombic $Pbca$ phase though the latter is favored thermodynamically. Additionally, the transition barriers depend strongly on the mechanical boundary conditions. In particular, when the shear deformation required for the tetragonal to monoclinic phase transition becomes inaccessible (e.g., clamping by grains or capping electrodes), the formation of the monoclinic phase will be suppressed whereas the formation of the polar orthorhombic phase is favored. 

Growing differently-oriented crystals on substates of unconventional cuts offers a new modality to tune the mechanical boundary conditions. We systematically investigate the effects of growth orientations and epitaxial strains on the relative stability of different phases of HfO$_2$. In agreement with previous studies, the biaxial compressive stress will drive the monoclinic to orthorhombic transition. Notably, the two orthorhombic phases at their optimal equilibrium strains are thermodynamically more stable than the monoclinic phase in (111)-oriented thin films. Combined with the kinetic effect that the transformation from the tetragonal phase to the polar orthorhombic phase is faster than to the non-polar orthorhombic phase, we propose it is likely to better stabilize the ferroelectric $Pca2_1$ phase in (111)-oriented thin films via appropriate thermal annealing process even in the absence of external electric fields. Recent realization of single orthorhombic phase in (111)-oriented ferroelectric Y-doped HfO$_2$ thin films further supports our findings. These results provide useful insights into the origin of ferroelectricity in HfO$_2$-based thin films and important implications for better control of the ferroelectric phase through substrate orientation engineering.

 \section{Acknowledgments}
SL is supported by SEDD Distinguished Postdoc Fellowship at US Army Research Laboratory.

\providecommand{\latin}[1]{#1}
\makeatletter
\providecommand{\doi}
  {\begingroup\let\do\@makeother\dospecials
  \catcode`\{=1 \catcode`\}=2 \doi@aux}
\providecommand{\doi@aux}[1]{\endgroup\texttt{#1}}
\makeatother
\providecommand*\mcitethebibliography{\thebibliography}
\csname @ifundefined\endcsname{endmcitethebibliography}
  {\let\endmcitethebibliography\endthebibliography}{}

\newpage
\begin{table}
\centering
\caption{Transition barriers between HfO$_2$ polymorphs. Energy in eV/f.u.}
\begin{tabular}{ccc}
\hline
\hline
  Phase Transition&Forward& Reverse\\
  \hline
T $\rightarrow$ AO & 0.067 & 0.162 \\
T$\rightarrow $ PO & 0.032 & 0.108\\
T $\rightarrow$ M  & 0.013& 0.141 \\
AO $\rightarrow$ PO & 0.171 & 0.151 \\
PO $\rightarrow$ M & 0.109 & 0.162\\
\hline
\hline
\end{tabular}
\end{table}

\newpage

\newpage
 
 \newpage
 
  \newpage
 
   \newpage
 
    \newpage
\end{document}